\documentclass[prd,twocolumn,nofootinbib]{revtex4}
\usepackage{dcolumn}
\usepackage{multirow}
\usepackage{graphicx}
\usepackage{amssymb}
\usepackage{bm}
\usepackage{hyperref}
\usepackage{epstopdf}
\usepackage{color}
\usepackage{mathrsfs}
\usepackage{amsmath,amssymb,amsthm}
\usepackage{rotating}
\usepackage{sverb, longtable}
\usepackage{subfigure}

\usepackage[graphicx]{realboxes}
\usepackage{adjustbox}

\newcommand{\Neff}{\ensuremath{N_{\rm eff}}}
\newcommand{\Neffstd}{\ensuremath{N_{\rm eff,std}}}
\begin{document}

\title{Scale and redshift  dependent limits on cosmic neutrino properties}
\author{Deng Wang}
\email{dengwang@ific.uv.es}
\author{Olga Mena}
\email{omena@ific.uv.es}
\affiliation{Instituto de F\'{i}sica Corpuscular (CSIC-Universitat de Val\`{e}ncia), E-46980 Paterna, Spain} 
\author{Eleonora Di Valentino}
\email{e.divalentino@sheffield.ac.uk}
\affiliation{School of Mathematical and Physical Sciences, University of Sheffield, Hounsfield Road, Sheffield S3 7RH, United Kingdom}
\author{Stefano Gariazzo}
\email{stefano.gariazzo@unito.it}
\affiliation{Department of Physics, University of Turin, via P.\ Giuria 1, 10125 Turin (TO), Italy \looseness=-1}
\affiliation{Istituto Nazionale di Fisica Nucleare (INFN), Sezione di Torino, via P.\ Giuria 1, 10125 Turin (TO), Italy}

\begin{abstract}
Cosmological neutrino mass and abundance measurements are reaching unprecedented precision. Testing their stability versus redshift and scale is a crucial issue, as it can serve as a guide for optimizing ongoing and future searches. Here, we perform such analyses, considering a number of redshift, scale, and redshift-and-scale nodes.
Concerning the $k$-space analysis of $\sum m_\nu$, CMB observations are crucial, as they lead the neutrino mass constraints. Interestingly, some data combinations suggest a non-zero value for the neutrino mass with $2\sigma$ significance. The most constraining bound we find is $\sum m_\nu<0.54$~eV at $95\%$~CL in the $[10^{-3}, 10^{-2}]$~$h$/Mpc $k$-bin, a limit that barely depends on the data combination. Regarding the redshift- and scale-dependent neutrino mass constraints, high redshifts ($z>100$) and scales in the range $[10^{-3}, 10^{-1}]$~$h$/Mpc provide the best constraints. The least constraining bounds are obtained at very low redshifts $[0,0.5]$ and also at very small scales ($k>0.1~h$/Mpc), due to the absence of observations. Highly relevant is the case of the $[100, 1100]$, $[10^{-2}, 10^{-1}]$~$h$/Mpc redshift-scale bin, where a $2$--$3\sigma$ evidence for a non-zero neutrino mass is obtained for all data combinations. The bound from CMB alone at $68\%$~CL is $0.63^{+0.20}_{-0.24}$~eV, and the one for the full dataset is $0.56^{+0.20}_{-0.23}$~eV, clearly suggesting a non-zero neutrino mass at these scales, possibly related to a deviation of the ISW amplitude in this redshift range.
Concerning the analysis of \Neff\ in the $k$-space, at intermediate scales ranging from $k=10^{-3}$~$h$/Mpc to $k=10^{-1}$~$h$/Mpc, accurate CMB data provide very strong bounds, the most robust one being \Neff$=3.09\pm 0.14$, comparable to the standard expected value without a $k$-bin analysis. If a non-zero neutrino mass is considered, the bounds on the \Neff\ values at the different $k$-bins are largely unaffected, and the $95\%$~CL tightest limit we find for the neutrino mass in this case is $\sum m_\nu < 0.205$~eV from the full dataset. Finally, the $z$ and $k$ analyses of \Neff\ indicate a high constraining power of cosmological observations at high redshifts and intermediate scales $[10^{-2}, 10^{-1}]$~$h$/Mpc when extracting the binned values of this parameter.

\end{abstract}
\maketitle

\section{Introduction}
Cosmological bounds on neutrino masses currently provide the tightest constraints on this fundamental quantity. Its precise value plays a key role not only from a purely theoretical perspective but also in the experimental roadmap, as neutrinoless double beta decay searches, which investigate the possible Majorana nature of neutrinos, depend on the absolute scale of neutrino masses and their hierarchical ordering. While neutrino oscillation experiments have measured two squared-mass differences, $|\Delta m^2_{31}|\equiv|m^2_3 - m^2_1| \simeq 2.5\cdot 10^{-3}\,\mathrm{eV}^2$ and $\Delta m^2_{21} \equiv m^2_2 - m^2_1 \simeq 7.5\cdot 10^{-5}\,\mathrm{eV}^2$~\cite{deSalas:2020pgw, Esteban:2024eli, Capozzi:2025wyn}, this leads to two possible mass orderings: \emph{normal} ($\Delta m^2_{31}>0$) and \emph{inverted} ($\Delta m^2_{31}<0$). 
A number of ongoing and future neutrino oscillation experiments aim to determine the neutrino mass ordering~\cite{T2K:2023smv, NOvA:2021nfi, JUNO:2015zny, Hyper-Kamiokande:2018ofw, DUNE:2020ypp}; see also Ref.~\cite{Gariazzo:2018pei}. Neutrino oscillation experiments provide a lower limit for the total neutrino mass, $\sum m_\nu > 0.06\,\mathrm{eV}$ for NO and $\sum m_\nu > 0.1\,\mathrm{eV}$ for IO~\cite{Esteban:2024eli}. Direct searches in the KATRIN experiment imply $\sum m_\nu < 1.35\,\mathrm{eV}$~\cite{Katrin:2024tvg}, where $\sum m_\nu \equiv m_1 + m_2 + m_3$ is the total neutrino mass.

The most constraining cosmological neutrino bound is approaching, or even surpassing, the minimum scale predicted by terrestrial oscillation searches, i.e., around $\sum m_\nu \lesssim 0.05$\,eV~\cite{Wang:2024hen}, leading to a tension between oscillation results and cosmological observations~\cite{Jiang:2024viw}.\footnote{Notice that cosmological inferences, however, are indirect. Relaxing the equation of state of dark energy~\cite{Hannestad:2005gj, Yang:2020ope, Vagnozzi:2018jhn, RoyChoudhury:2018vnm, diValentino:2022njd, Zhao:2016ecj, Zhang:2015uhk, Guo:2018gyo, RoyChoudhury:2019hls, Li:2012vn, Li:2012spm, Zhang:2014nta, Zhang:2015rha, Geng:2015haa, Chen:2015oga, Loureiro:2018pdz, Wang:2016tsz, Yang:2017amu, Huang:2015wrx, Sharma:2022ifr, Zhang:2020mox, Khalifeh:2021ree, Chen:2016eyp}, the Hubble constant $H_0$~\cite{Planck:2018vyg, Giusarma:2012ph, Hu:2023jqc, Schoneberg:2021qvd}, the amplitude parameter $\sigma_8$~\cite{Planck:2018vyg, Mccarthy:2017yqf, Abazajian:2019ejt}, which are in tension among different datasets~\cite{Abdalla:2022yfr, DiValentino:2021izs, Verde:2019ivm, Verde:2023lmm, DiValentino:2020vvd, RoyChoudhury:2018gay}; CMB lensing~\cite{RoyChoudhury:2019hls, DiValentino:2021imh, Giare:2023aix, Craig:2024tky, Green:2024xbb, Loverde:2024nfi, Naredo-Tuero:2024sgf, Allali:2024aiv}, or allowing new physics in the neutrino sector~\cite{Esteban:2021ozz, Esteban:2022rjk, Kreisch:2019yzn, Lattanzi:2017ubx, Chacko:2019nej, Chacko:2020hmh, Escudero:2020ped, FrancoAbellan:2021hdb, Bellomo:2016xhl, Dvali:2016uhn, Lorenz:2018fzb, Dvali:2021uvk, Lorenz:2021alz, Escudero:2022gez, Sen:2023uga, Farzan:2015pca, Alvey:2021xmq, Oldengott:2019lke, Alvey:2021sji, Cuoco:2005qr, Allali:2024anb, Benso:2024qrg} or isolating background from perturbation effects~\cite{Bertolez-Martinez:2024wez} may relax the limit quoted before.}

On the other hand, the number of neutrino species is commonly parameterized in terms of the effective number of relativistic degrees of freedom, \Neff, by
\begin{equation}
\rho_{\rm R} = \rho_\gamma \left( 1 + \frac{7}{8} \left(\frac{4}{11}\right)^{4/3} \Neff \right)\,,
\end{equation}
where $\rho_\gamma$ represents the photon energy density. Here, \Neff\ includes the contribution of all relativistic particles besides photons. If we consider the simplest three-neutrino case with an instantaneous decoupling process, \Neff\ would be equal to 3. A value $\Neff \neq 3$, however, would be the consequence of either new degrees of freedom unrelated to standard neutrinos, a non-standard momentum distribution for the three neutrinos, or other exotic new physics in the neutrino sector, such as the presence of additional neutrino states~\citep{Gariazzo:2019gyi, Gariazzo:2022evs}. 
Even in the standard case, the value of \Neff\ deviates from 3 due to non-instantaneous decoupling. This deviation can be computed numerically by considering the full framework of neutrino oscillations, interactions with electrons and positrons, Finite-Temperature corrections to Quantum Electro-Dynamics (FT-QED), and the expansion of the universe, leading to $\Neffstd = 3.044$~\citep{Akita:2020szl, Froustey:2020mcq, Bennett:2020zkv, Cielo:2023bqp, Drewes:2024wbw}, see also~\citep{Mangano:2005cc, deSalas:2016ztq}. 
Current cosmological observations indicate that \Neff\ is close to 3, as measured independently by CMB observations ($\Neff=2.99^{+0.34}_{-0.33}$ at 95\% confidence level (CL)~\citep{Planck:2018vyg}) and BBN abundances (e.g., $\Neff=2.87^{+0.24}_{-0.21}$ at 68\% CL~\citep{Consiglio:2017pot}). The above constraints have been shown to be extremely robust against different fiducial cosmologies; see Ref.~\citep{diValentino:2022njd}.\footnote{The largest departure concerning the uncertainties in \Neff\ appears in models that also consider the Helium fraction as a free parameter, due to the degeneracy between \Neff\ and $Y_p^{\rm BBN}$ via the Silk damping effect.}
Given the fact that neutrinos transition from a purely relativistic radiation component in the early Universe to non-relativistic species (contributing to the Universe's matter energy density) at late times, it is mandatory to test whether the cosmological measurements of neutrino masses and abundances hold throughout cosmic evolution and to assess the capability of different datasets to constrain both properties at various cosmological scales and times. 
Redshift-dependent cosmological limits on neutrino masses and abundances have been computed previously in the literature~\cite{Lorenz:2021alz,Safi:2024bta}. Here, we go beyond these studies by including an analysis of the stability of cosmological neutrino limits with respect to both \emph{scale} and the combination of \emph{redshift and scale}. We conduct these analyses to gain insight into the constraining power of cosmological datasets regarding neutrino properties, not only as a function of time but also as a function of scale. This will be invaluable for identifying which scales are most sensitive to, e.g., the neutrino mass—crucial information for future cosmological measurements, which are expected to detect the minimum neutrino mass and determine \Neff\ with unprecedented precision~\cite{Euclid:2024imf, DESI:2016fyo, SimonsObservatory:2018koc, LiteBIRD:2022cnt}.
While the aim of this manuscript is to perform a pure phenomenological analysis of the potential redshift and/or scale dependence of the neutrino parameters, there are a number of theoretical frameworks that could potentially lead to such scenarios. Neutrino decays~\cite{Escudero:2020ped,Chacko:2020hmh,Chacko:2019nej}, long-range neutrino forces~\cite{Esteban:2021ozz}, neutrino cooling and heating~\cite{Craig:2024tky}, mass-varying neutrino models~\cite{Fardon:2003eh}, among others~\cite{Dvali:2016uhn}, can benefit from the analyses presented here.

The structure of this manuscript is as follows. Section~\ref{sec:Observational data sets} contains a description of the datasets employed in this work. Section~\ref{sec:modeling} outlines the redshift and scale binning adopted in the numerical analyses, presented in terms of scale, redshift, and the combination of both for neutrino masses and abundances. Section~\ref{sec:method} details the methodology adopted. We present our results in Sec.~\ref{sec:results}, and finally, we conclude in Sec.~\ref{sec:conclusions}.

\section{Data, modeling and methodology}
\subsection{Observational data sets}
\label{sec:Observational data sets}

In the following, we describe the datasets used to perform our analysis.
\begin{itemize}
    \item \textbf{CMB.} We consider the Planck 2018 high-$\ell$ \texttt{plik} temperature (TT) likelihood covering multipoles $30\leqslant\ell\leqslant2508$, polarization (EE), and their cross-correlation (TE) data spanning $30\leqslant\ell\leqslant1996$. Furthermore, we incorporate the low-$\ell$ TT \texttt{Commander} and \texttt{SimAll} EE likelihoods in the range $2\leqslant\ell\leqslant29$~\cite{Planck:2019nip}. In addition, we conservatively include the Planck CMB lensing likelihood derived from \texttt{SMICA} maps across $8\leqslant\ell \leqslant400$~\cite{Planck:2018lbu}, namely TTTEEE+low$\ell$+lowE+lensing~\cite{Planck:2018vyg,Planck:2019nip,Planck:2018nkj,planck}. The Planck datasets are expected to provide constraints covering scales $10^{-4} \lesssim k \, [h\,\mathrm{Mpc}^{-1}] \lesssim 0.3$. Moreover, the CMB originated at redshift $z \simeq 1100$, but the evolution of the Universe has an impact on its spectrum even at much later times. As a consequence, the CMB dataset can probe redshifts down to $z = \mathcal{O}(10)$.
    An upgraded version of the likelihood for Planck data has been released, providing very similar results to those obtained with previous likelihoods, albeit with $10\%$ to $20\%$ smaller uncertainties in the extraction of the cosmological parameters within the $\Lambda$CDM model~\cite{Tristram:2023haj}. We shall consider this new and improved likelihood in future analyses.
    
    \item \textbf{MPS.} The galaxy power spectrum is an important quantity for measuring the large-scale structure of the Universe at late times. It can be transformed into the matter power spectrum (MPS) by modeling the so-called galaxy bias. In this analysis, we use the galaxy power spectrum data measured at four effective redshifts, $z_{\rm eff} = 0.22$, 0.41, 0.60, and 0.78, from the WiggleZ Dark Energy Survey~\cite{Blake:2010xz,Parkinson:2012vd}, which measured 238,000 galaxy redshifts from seven regions of the sky with a total volume of 1 Gpc$^{3}$ in the scale range $[0.01,\,0.5]$~$h$ Mpc$^{-1}$.

    \item \textbf{DESY1.} The Dark Energy Survey Year 1 (DESY1) large-scale structure observations~\cite{DES:2017myr,DES:2017gwu,DES:2017hdw,DES:2017qwj,DES:2018ufa} include the following three two-point correlation functions: 
    
    \begin{enumerate}
        \item \textit{Galaxy clustering.} The homogeneity of the matter distribution in the Universe can be traced by the distribution of galaxies. The overabundance of galaxy pairs at an angular separation $\theta$ in a random distribution, $\omega(\theta)$, is one of the most convenient ways to measure galaxy clustering. It quantifies the scale dependence and strength of galaxy clustering and, consequently, affects matter clustering. 
        
        \item \textit{Cosmic shear.} The two-point statistics describing galaxy shapes are complex since they result from components of a spin-2 tensor. Thus, it is convenient to extract information from a galaxy survey using a pair of two-point correlation functions, $\xi_+(\theta)$ and $\xi_-(\theta)$, which represent the sum and difference of products of the tangential and cross components of the shear, measured with respect to the line connecting each pair of galaxies. 
        
        \item \textit{Galaxy-galaxy lensing.} The typical distortion of source galaxy shapes arises from masses associated with foreground lenses. This distortion is characterized by the mean tangential ellipticity of source galaxy shapes around lens galaxy positions for each pair of redshift bins and is consequently referred to as the tangential shear, $\zeta_t(\theta)$. 
    \end{enumerate}
    
    DESY1 observations cover the range $0.15 \lesssim z \lesssim 0.9$ for the galaxy samples and $0.2 \lesssim z \lesssim 1.3$ for cosmic shear. Concerning scales, the tested range is approximately $[0.01,\,0.5]$~$h$ Mpc$^{-1}$.
    
\end{itemize}

\subsection{Scale and redshift dependent model}
\label{sec:modeling} 
Concerning the \emph{scale-dependent analyses}, we make use of seven $k$-bins for $\sum m_\nu$ and \Neff, namely: $[0, 10^{-4}]$, $[10^{-4}, 10^{-3}]$, $[10^{-3}, 10^{-2}]$, $[10^{-2}, 10^{-1}]$, $[10^{-1}, 1]$, $[1, 10]$, and $[10, +\infty)$, all in $h$/Mpc units. The parameter values at these nodes are 
{\boldmath$\Sigma m_\nu^{k7}$}, {\boldmath$\Sigma m_\nu^{k6}$}, {\boldmath$\Sigma m_\nu^{k5}$}, {\boldmath$\Sigma m_\nu^{k4}$}, {\boldmath$\Sigma m_\nu^{k3}$}, 
{\boldmath$\Sigma m_\nu^{k2}$}, {\boldmath$\Sigma m_\nu^{k1}$}, and  
{\boldmath$N_{\rm eff}^{k7}$}, {\boldmath$N_{\rm eff}^{k6}$}, {\boldmath$N_{\rm eff}^{k5}$}, {\boldmath$N_{\rm eff}^{k4}$}, {\boldmath$N_{\rm eff}^{k3}$}, {\boldmath$N_{\rm eff}^{k2}$}, {\boldmath$N_{\rm eff}^{k1}$}.

Regarding the scale- and redshift-dependent constraints for $\sum m_\nu$, we adopt six redshift bins ($i$): $[0, 0.5]$, $[0.5, 3]$, $[3, 10]$, $[10, 100]$, $[100, 1100]$, and $[1100, +\infty)$, along with four bins in $k$-space ($j$): $[10^{-1}, +\infty)$, $[10^{-2}, 10^{-1}]$, $[10^{-3}, 10^{-2}]$, and $[0, 10^{-3}]$ (all in $h$/Mpc units). The values of $\sum m_\nu$ at these nodes are {\boldmath$\Sigma m_\nu^{ij}$}, with $i=1,\dots,6$ and $j=1,\dots,4$, leading to a total of 24 additional parameters. 

Given that \Neff\ is primarily constrained by recombination physics, we consider only three redshift bins: $[0, 100]$, $[100, 1100]$, and $[1100, +\infty)$, along with three $k$-space bins: $[10^{-1}, +\infty)$~$h$/Mpc, $[10^{-2}, 10^{-1}]$~$h$/Mpc, and $[0, 10^{-2}]$~$h$/Mpc. The values of \Neff\ at these nodes are {\boldmath$\Neff^{ij}$}, with $i=1,\dots,3$ and $j=1,\dots,3$, resulting in a total of 9 new parameters.

In practice, we do not directly modify background and perturbation equations, but add several subroutines in \texttt{CAMB}~\cite{Lewis:1999bs}, where the above-mentioned tomographic method are well implemented, to evolve the background and perturbed energy densities and heat flux of massive neutrinos over redshifts and/or scales. All the tomographic parameters are included in these subroutines. These subroutines can control which redshifts and/or scales are included and consequently help explore the neutrino mass limits through the cosmic history and multiple scales.

In all the models considered here for neutrino masses (number of relativistic species), we assume $\sum m_\nu$ $(N_{\rm eff})$ to be a constant in each bin. Specifically, for the case of scale-dependent $\sum m_\nu$ and scale- and redshift-dependent $\sum m_\nu$, setting $N_{\rm eff}=3.044$, we take $\sum m_\nu=0$ eV at the background level and pass the perturbed energy density and heat flux of massive neutrinos to the corresponding perturbation equations implemented in \texttt{CAMB} in each bin with the degenerate neutrino hierarchy. Consequently, distance probes cannot influence the constraints on the scale-related $\sum m_\nu$ parameters in such settings. For the scale-dependent $N_{\rm eff}$ and scale- and redshift-dependent $N_{\rm eff}$, we adopt $N_{\rm eff}=3.044$ at the background level and, similarly, pass the perturbed energy density, heat flux and anisotropic stress ($\sigma_\nu$) of massless neutrinos to the corresponding perturbation equations, while assuming $\sum m_\nu=0$ eV at both the background and the perturbation levels. However, for simplicity, we assume $\sigma_\nu=0$ in our numerical analyses.

\subsection{Analysis methodology}\label{sec:method}

As we have discussed, in order to calculate the theoretical power spectra and the background evolution of the Universe, we use a modified version of the Boltzmann solver \texttt{CAMB}~\cite{Lewis:1999bs}. For the Bayesian analysis, we employ the Monte Carlo Markov Chain (MCMC) method to infer the posterior distributions of model parameters, leveraging the publicly available package \texttt{CosmoMC}~\cite{Lewis:2002ah,Lewis:2013hha}. We assess the convergence of the MCMC chains via the Gelman-Rubin statistic criterion, requiring $R - 1 \lesssim 0.05$~\cite{Gelman:1992zz}, and analyze the chains using the package \texttt{Getdist}~\cite{Lewis:2019x}.

We use the following uniform priors for the model parameters: the baryon fraction $\Omega_b h^2 \in [0.005, 0.1]$, the cold dark matter fraction $\Omega_c h^2 \in [0.001, 0.99]$, the acoustic angular scale at the recombination epoch $100\theta_{MC} \in [0.5, 10]$, the scalar spectral index $n_s \in [0.8, 1.2]$, the amplitude of the primordial scalar power spectrum $\ln(10^{10}A_s) \in [2, 4]$, the optical depth $\tau \in [0.01, 0.8]$, and the sum of the masses of the three active neutrinos $\sum m_\nu \in [0, 5]$~eV.
For the scale-dependent neutrino mass model, we use the prior {\boldmath$\sum m_\nu^{k_i}$} $\in [0, 30]$~eV with $i=1,\dots,7$ in each $k$-bin. For the scale- and redshift-dependent neutrino mass model, we use the prior {\boldmath$\sum m_\nu^{ij}$} $\in [0, 30]$~eV with $i=1,\dots,6$ and $j=1,\dots,4$ in each bin.
For the scale-dependent number of relativistic species model, both without and with massive neutrinos, we use the prior {\boldmath$N_{\rm eff}^{k_i}$} $\in [0, 30]$ with $i=1,\dots,7$ in each $k$-bin.  
For the scale- and redshift-dependent number of relativistic species model, we use the prior {\boldmath$N_{\rm eff}^{ij}$} $\in [0, 30]$ with $i=1,\dots,3$ and $j=1,\dots,3$ in each bin.

\section{Results}
\label{sec:results} 

\subsection{Scale and redshift dependent neutrino mass limits}
\begin{figure}
	\centering
	\includegraphics[scale=0.55]{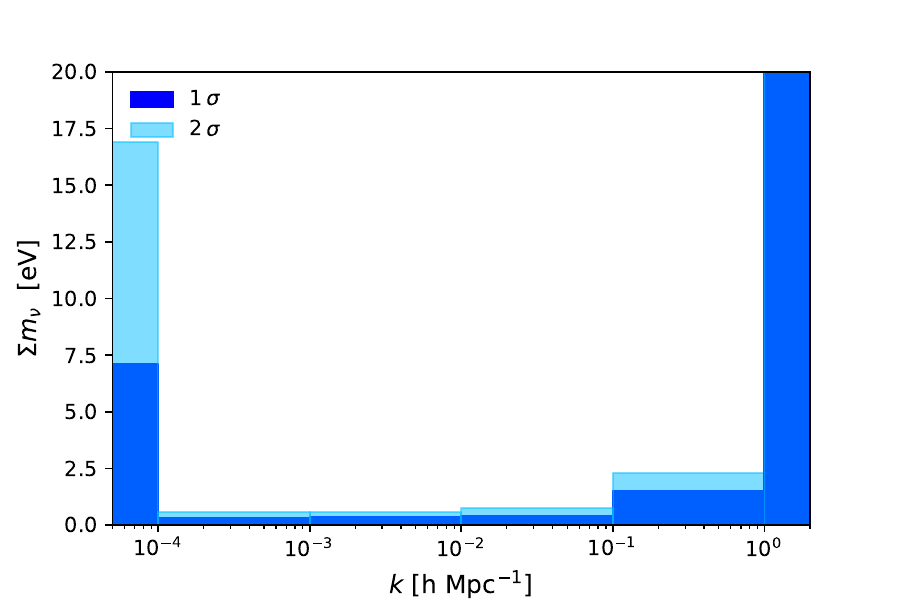}
	\caption{The tomographic reconstruction of neutrino masses across cosmic scales from CMB data.}\label{fig:mnukevolution}
	
\end{figure}

\begin{figure*}
	\centering
	\includegraphics[scale=0.5]{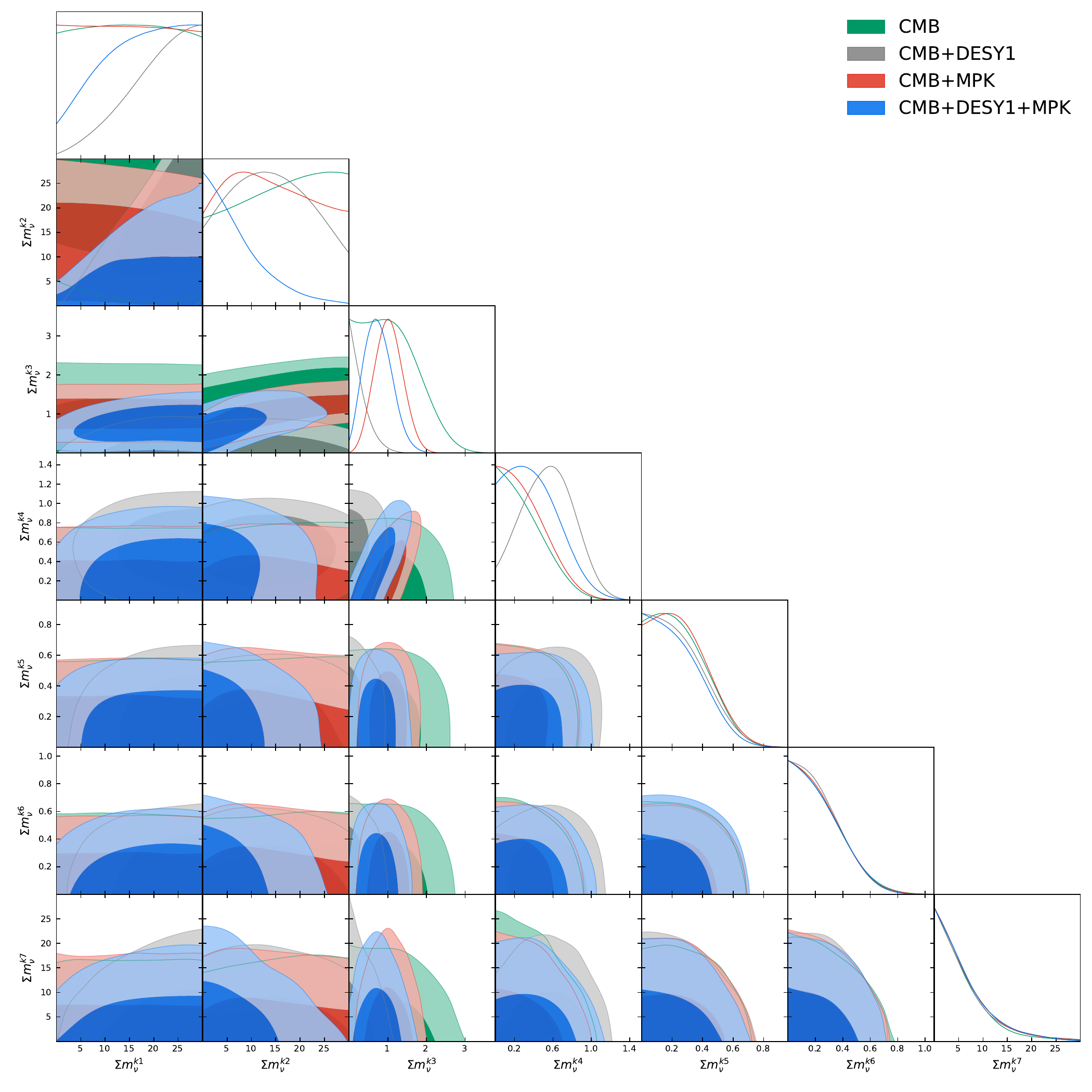}
	\caption{One-dimensional and two-dimensional marginalized constraints on the scale-dependent neutrino mass model from CMB, CMB plus DESY1, CMB plus MPK and CMB plus DESY1 plus MPK datasets, respectively.}\label{fig:mnuk}
	
\end{figure*}

Table~\ref{fig:mnukevolution} shows the constraints on the main cosmological parameters, as well as on the different values of $\sum m_\nu$ at the corresponding bins in $k$-space. Notice that {\boldmath$\Sigma m_\nu^{k7}$}, corresponding to very large scales, is poorly constrained regardless of the datasets used in the analyses. The lack of observational measurements at these very large scales leads to a very loose bound on this parameter. The situation changes completely when moving to the next two $k$-bins, i.e., the $[10^{-4}, 10^{-3}]$~$h$/Mpc and the $[10^{-3}, 10^{-2}]$~$h$/Mpc bins.
Here, CMB observations are crucial, as they dominate the neutrino mass constraints, as can also be inferred from Figs.~\ref{fig:mnukevolution} and~\ref{fig:mnuk}: the additional datasets related to large-scale structure data do not further improve the bounds on either {\boldmath$\Sigma m_\nu^{k6}$} or {\boldmath$\Sigma m_\nu^{k5}$}.
The following two bins are particularly interesting, as some data combinations suggest a non-zero value for the neutrino mass. More concretely, within the $[10^{-2}, 10^{-1}]$~$h$/Mpc $k$-bin, the CMB plus DESY1 datasets together provide the bound {\boldmath$\Sigma m_\nu^{k5}$}$ = 0.55 \pm 0.26$~eV with $68\%$~CL errors, indicating a non-zero neutrino mass at the $2\sigma$ significance level (see also Fig.~\ref{fig:mnuk}). For the rest of the data combinations, the limit on {\boldmath$\Sigma m_\nu^{k5}$} remains very similar to the CMB-only result.
For the {\boldmath$\Sigma m_\nu^{k4}$} parameter, associated with the $[10^{-1}, 1]$~$h$/Mpc $k$-bin, the situation is different: in general, the addition of large-scale structure data significantly alters the CMB-only limit, as these scales play a leading role in constraining neutrino mass. For instance, the addition of DESY1 measurements to CMB changes the $95\%$~CL upper bound on {\boldmath$\Sigma m_\nu^{k3}$} from $2.30$~eV to $0.82$~eV. When adding MPS observations to CMB, we find $\sim 3\sigma$ evidence for a non-zero neutrino mass, {\boldmath$\Sigma m_\nu^{k3}$}$ = 1.02 \pm 0.36$~eV. For the full dataset combination (CMB plus DESY1 plus MPS), we obtain {\boldmath$\Sigma m_\nu^{k3}$}$ = 0.75^{+0.31}_{-0.41}$~eV, suggesting a non-zero value for the neutrino mass with $2\sigma$ significance.
Within the remaining bins in $k$-space, corresponding to very small scales, the neutrino mass is largely unconstrained, except in a few cases where DESY1 data are considered along with other datasets. Specifically, DESY1 marginally contributes in these cases. For instance, we obtain {\boldmath$\Sigma m_\nu^{k2}$}$ < 18.70$~eV at $95\%$~CL for the $[1, 10]$~$h$/Mpc $k$-bin.

The redshift- and scale-dependent neutrino mass constraints are depicted in Tab.~\ref{tab:mnuzk}. Notice that, at the highest redshift bin, $z > 1100$, the CMB constraining power on neutrino masses barely changes when considering additional datasets, which cannot probe such high redshift values. Only in the $k$-range $[10^{-2}, 10^{-1}]$~$h$/Mpc are these additional measurements mildly significant: for instance, the $95\%$~CL upper limit on {\boldmath$\Sigma m_\nu^{62}$} of $0.85$~eV from CMB data alone is shifted down to $0.79$~eV when also including the MPS dataset. For smaller scales, the bound on {\boldmath$\Sigma m_\nu^{61}$} of $3.50$~eV from CMB data alone remains unchanged when incorporating large-scale structure constraints, due to the lack of precise large-scale structure measurements in the $[10^{-1}, +\infty)$~$h$/Mpc $k$-bin, where non-linearities are crucial. 

The $[100, 1100]$ redshift range also benefits from large-scale structure measurements, especially in the $[10^{-3}, 10^{-2}]$~$h$/Mpc and $[10^{-2}, 10^{-1}]$~$h$/Mpc $k$-bins. Highly relevant is the case of {\boldmath$\Sigma m_\nu^{52}$}, where a $2$--$2.5\sigma$ evidence for a non-zero neutrino mass is obtained for all data combinations. The bound from CMB alone is $0.63^{+0.20}_{-0.24}$~eV, while the bound for the full dataset combination (CMB plus DESY1 plus MPS) is $0.56^{+0.20}_{-0.23}$~eV, clearly suggesting a non-zero neutrino mass at these scales within the $[100, 1100]$ redshift range. 
This signal could be related to the fact that there is a $\sim 2\sigma$ deviation of the ISW amplitude ($A_{\textrm{ISW}} > 1$) at redshift $z = 500$, as indicated by the tomographic analyses of Ref.~\cite{Wang:2024kpu}. Since neutrino masses reduce the ISW effect, a relatively large neutrino mass (compared to the current cosmological limits) could, in principle, compensate for such an effect.
   
For the $[10, 100]$ redshift bin, the tightest constraints are found within the $[10^{-3}, 10^{-2}]$~$h$/Mpc $k$-bin, where CMB measurements are highly relevant and dominant: the CMB-only limit of {\boldmath$\Sigma m_\nu^{43}$}$ < 1.11$~eV at $95\%$~CL does not change when additional measurements are considered. At the smallest scales, the neutrino mass is unconstrained—i.e., no limit on {\boldmath$\Sigma m_\nu^{41}$} is found—while the limits on {\boldmath$\Sigma m_\nu^{42}$} are also very weak. 
The situation in the following redshift bin, $[3, 10]$, is very similar, although in this case the CMB-only limit is higher: {\boldmath$\Sigma m_\nu^{33}$}$ < 3.10$~eV at $95\%$~CL.
It is precisely in the $[0.5, 3]$ redshift range where the effects of large-scale structure data become relevant, especially within the $[10^{-2}, 10^{-1}]$~$h$/Mpc $k$-bin. The $95\%$~CL limit on {\boldmath$\Sigma m_\nu^{22}$}$ < 5.90$~eV from CMB data alone is reduced to $2.20^{+1.00}_{-1.50}$~eV when the full dataset combination is considered. Even more interesting is the case of {\boldmath$\Sigma m_\nu^{21}$}, which is completely unconstrained by CMB observations but gains constraints at $95\%$~CL of $3.20$~eV and $5.00$~eV when adding DESY1 and MPS measurements to CMB data, respectively. When considering the full dataset, we obtain {\boldmath$\Sigma m_\nu^{21}$}$ = 2.15^{+0.95}_{-1.40}$~eV, indicating a very mild preference for a non-zero neutrino mass.
The lowest redshift range provides very weak constraints on the neutrino mass, especially at the highest and lowest scales. Only the value of {\boldmath$\Sigma m_\nu^{13}$} is always constrained, albeit very poorly. The lack of observations in these ranges leads to very weak bounds on the neutrino mass.

\begin{figure*}
	\centering
	\includegraphics[scale=0.55]{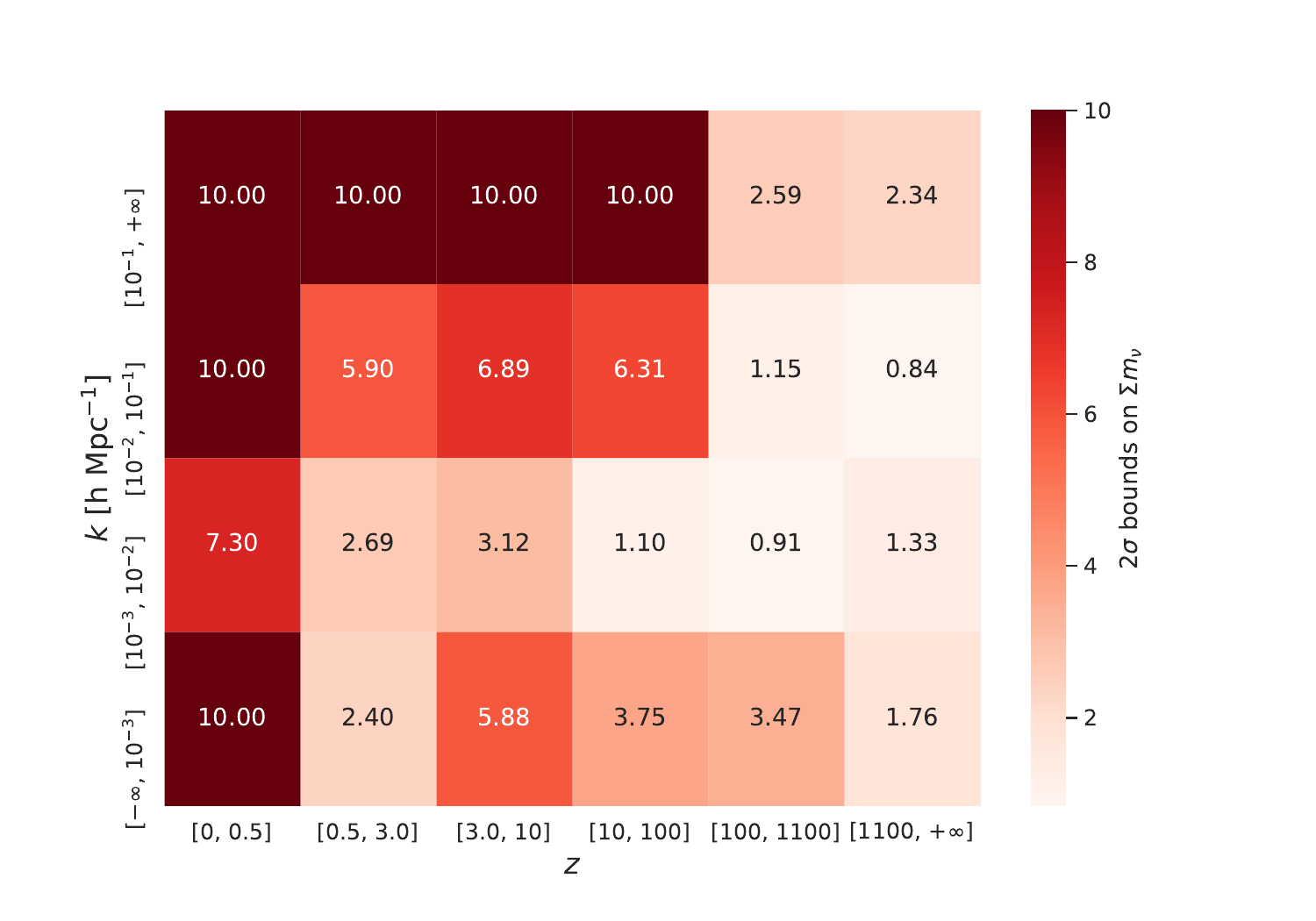}
	\caption{$2\sigma$ bounds on neutrino masses in the redshift- and scale-dependent neutrino mass model from CMB data. Note that all bins have $2\sigma$ upper bounds except for {\boldmath$\Sigma m_\nu^{64}$}, which has a $2\sigma$ lower limit of 1.76 eV. The value ``10.00'' denotes parameters that remain unconstrained by the data. }\label{fig:mnuzkmap}
	
\end{figure*}

Figure~\ref{fig:mnuzkmap} depicts the results for the neutrino mass constraints as a function of redshift and $k$-bins. Notice that high redshifts ($z > 100$) and scales in the range $[10^{-3}, 10^{-1}]$~$h$/Mpc provide the best constraints. The least constraining bounds are obtained at very low redshifts, $[0, 0.5]$, and at very small scales ($k > 0.1~h$/Mpc), due to the absence of observations. In this particular region of $k$-space, a bound on the neutrino mass is found only if $z > 100$.

\begin{table*}[!t]
	\renewcommand\arraystretch{1.6}
	\caption{Mean values and $1\,\sigma$ ($68\%$) marginalized errors of the cosmological parameters within the scale-dependent neutrino mass model, obtained using the CMB, CMB plus DESY1, CMB plus MPS, and CMB plus DESY1 plus MPS datasets, respectively. Note that we quote $2\,\sigma$ ($95\%$~CL) limits for parameters that cannot be measured by the data. The symbols ``$\bigstar$'' denote parameters that remain unconstrained by the data.}
	\setlength{\tabcolsep}{5mm}{
		\begin{tabular} { l | c | c | c | c }
			\hline
			\hline
			Parameters              &  CMB & CMB+DESY1 & CMB+MPS & CMB+DESY1+MPS    \\
			\hline
			{\boldmath$\Omega_b h^2   $} & $0.02242\pm 0.00015        $ & $0.02249\pm 0.00014        $ & $0.02240\pm 0.00014        $ & $0.02246\pm 0.00014        $\\

{\boldmath$\Omega_c h^2   $} & $0.1196\pm 0.0012          $ & $0.1186\pm 0.0011          $ & $0.1199\pm 0.0011          $ & $0.1189^{+0.0011}_{-0.00099}$\\

{\boldmath$100\theta_{MC} $} & $1.04091\pm 0.00031        $ & $1.04100\pm 0.00031        $ & $1.04087\pm 0.00031        $ & $1.04098\pm 0.00030        $\\

{\boldmath$\tau           $} & $0.0545\pm 0.0075          $ & $0.0550\pm 0.0076          $ & $0.0530\pm 0.0073          $ & $0.0543\pm 0.0071          $\\

{\boldmath${\rm{ln}}(10^{10} A_s)$} & $3.045\pm 0.015            $ & $3.044\pm 0.015            $ & $3.042\pm 0.014            $ & $3.043\pm 0.014            $\\

{\boldmath$n_s            $} & $0.9663\pm 0.0044          $ & $0.9668\pm 0.0042          $ & $0.9651\pm 0.0040          $ & $0.9665\pm 0.0040          $\\

{\boldmath$\Sigma m_\nu^{k1}         $} & $\bigstar$                          & $> 7.23                    $ & $\bigstar$                          & $\bigstar$                        \\

{\boldmath$\Sigma m_\nu^{k2}         $} & $\bigstar$                         & $\bigstar$                          & $\bigstar$                         & $< 19.9                    $\\

{\boldmath$\Sigma m_\nu^{k3}         $} & $< 2.28                    $ & $< 0.806                   $ & $1.02\pm 0.36      $ & $0.75^{+0.31}_{-0.41}      $\\

{\boldmath$\Sigma m_\nu^{k4}         $} & $< 0.744                   $ & $0.55\pm 0.26      $ & $< 0.761                   $ & $< 0.875                   $\\

{\boldmath$\Sigma m_\nu^{k5}         $} & $< 0.563                   $ & $< 0.570                   $ & $< 0.574                   $ & $< 0.542                   $\\

{\boldmath$\Sigma m_\nu^{k6}         $} & $< 0.572                   $ & $< 0.552                   $ & $< 0.563                   $ & $< 0.560                   $\\

{\boldmath$\Sigma m_\nu^{k7}         $} & $< 16.5                    $ & $< 18.1                    $ & $< 17.7                    $ & $< 17.6                    $\\

\hline

{\boldmath $H_0   $} & $67.60\pm 0.55             $ & $68.06\pm 0.51             $ & $67.45\pm 0.49             $ & $67.90^{+0.45}_{-0.50}     $\\

{\boldmath $\Omega_m    $} & $0.3123\pm 0.0075          $ & $0.3060\pm 0.0067          $ & $0.3144\pm 0.0067          $ & $0.3082\pm 0.0064          $\\

{\boldmath $\sigma_8                  $} & $0.771^{+0.025}_{-0.022}   $ & $0.787^{+0.015}_{-0.010}   $ & $0.771^{+0.018}_{-0.016}   $ & $0.775^{+0.020}_{-0.016}   $\\

{\boldmath$S_8                       $} & $0.787\pm 0.025            $ & $0.794^{+0.017}_{-0.014}   $ & $0.789\pm 0.020            $ & $0.785^{+0.021}_{-0.019}   $\\

			\hline
			\hline
		\end{tabular}
		\label{tab:mnuk}}
\end{table*}

\begin{table*}[!t]
	\renewcommand\arraystretch{1.6}
	\caption{Mean values and $1\,\sigma$ ($68\%$) marginalized errors of the cosmological parameters within the redshift- and scale-dependent neutrino mass model, obtained using the CMB, CMB plus DESY1, CMB plus MPS and CMB plus DESY1 plus MPS datasets, respectively. Note that we quote $2\,\sigma$ ($95\%$~CL) limits for parameters that cannot be measured by the data. The symbols ``$\bigstar$'' denote parameters that remain unconstrained by the data.}
	\setlength{\tabcolsep}{5mm}{
		\begin{tabular} { l |c| c |c| c  }
			\hline
			\hline
			Parameters              &  CMB & CMB+DESY1 & CMB+MPS & CMB+DESY1+MPS    \\
			\hline
			{\boldmath$\Omega_b h^2   $} & $0.02245\pm 0.00019        $ & $0.02253\pm 0.00018        $ & $0.02245\pm 0.00019        $ & $0.02254\pm 0.00018        $\\

                {\boldmath$\Omega_c h^2   $} & $0.1226\pm 0.0016          $ & $0.1207\pm 0.0014          $ & $0.1226\pm 0.0015          $ & $0.1209\pm 0.0013          $\\

                {\boldmath$100\theta_{MC} $} & $1.04063\pm 0.00032        $ & $1.04081\pm 0.00032        $ & $1.04064\pm 0.00032        $ & $1.04081\pm 0.00031        $\\

                {\boldmath$\tau           $} & $0.0577^{+0.0077}_{-0.0087}$ & $0.0603\pm 0.0082          $ & $0.0571\pm 0.0083          $ & $0.0602^{+0.0074}_{-0.0083}$\\
			
			{\boldmath${\rm{ln}}(10^{10} A_s)$} & $3.061\pm 0.016            $ & $3.062\pm 0.016            $ & $3.059\pm 0.016            $ & $3.061^{+0.014}_{-0.016}   $\\

                {\boldmath$n_s            $} & $0.9574\pm 0.0059          $ & $0.9613\pm 0.0057          $ & $0.9579\pm 0.0057          $ & $0.9619\pm 0.0052          $\\

{\boldmath$\Sigma m_\nu^{11}$} & $\bigstar$                          & $6.40^{+2.80}_{-1.60}         $ & $3.70^{+1.20}_{-3.10}         $ & $5.1\pm 2.2        $\\

{\boldmath$\Sigma m_\nu^{12}$} & $\bigstar$                          & $6.10^{+2.20}_{-1.70}         $ & $<8.30         $ & $4.9\pm 2.3         $\\

{\boldmath$\Sigma m_\nu^{13}$} & $<7.30                    $ & $<5.10         $ & $<8.20        $ & $<6.70         $\\

{\boldmath$\Sigma m_\nu^{14}$} & $\bigstar$                          & $\bigstar        $ & $\bigstar         $ & $\bigstar         $\\

{\boldmath$\Sigma m_\nu^{21}$} & $\bigstar$                          & $<3.20         $ & $<5.00         $ & $2.15^{+0.95}_{-1.40}         $\\

{\boldmath$\Sigma m_\nu^{22}$} & $< 5.90                    $ & $2.42^{+0.86}_{-1.20}          $ & $<4.90         $ & $2.20^{+1.00}_{-1.50}         $\\

{\boldmath$\Sigma m_\nu^{23}$} & $< 2.69                    $ & $<2.30         $ & $<2.80         $ & $<2.50         $\\

{\boldmath$\Sigma m_\nu^{24}$} & $< 2.40                    $ & $<2.45       $ & $<2.53       $ & $<2.50         $\\

{\boldmath$\Sigma m_\nu^{31}$} & $\bigstar$                 & $\bigstar         $ & $\bigstar         $ & $\bigstar         $\\

{\boldmath$\Sigma m_\nu^{32}$} & $< 6.89                    $ & $<5.00         $ & $<7.10         $ & $<5.20         $\\

{\boldmath$\Sigma m_\nu^{33}$} & $< 3.12                    $ & $<3.20         $ & $<3.20         $ & $<3.30         $\\

{\boldmath$\Sigma m_\nu^{34}$} & $< 5.88                    $ & $3.10^{+1.50}_{-1.80}         $ & $<6.00         $ & $<5.90         $\\

{\boldmath$\Sigma m_\nu^{41}$} & $\bigstar$                  & $\bigstar          $ & $\bigstar          $ & $4.7^{+5.0}_{-4.5}         $\\

{\boldmath$\Sigma m_\nu^{42}$} & $< 6.31                    $ & $<5.60         $ & $<6.40         $ & $<5.60         $\\

{\boldmath$\Sigma m_\nu^{43}$} & $< 1.10                    $ & $<1.23      $ & $<1.16      $ & $<1.09      $\\

{\boldmath$\Sigma m_\nu^{44}$} & $< 3.75                    $ & $<3.60         $ & $<3.70         $ & $<3.50         $\\

{\boldmath$\Sigma m_\nu^{51}$} & $< 2.59                    $ & $<2.70         $ & $<2.90         $ & $<2.70         $\\

{\boldmath$\Sigma m_\nu^{52}$} & $0.63^{+0.20}_{-0.24}      $ & $0.57^{+0.22}_{-0.26}      $ & $0.61\pm 0.23              $ & $0.56^{+0.20}_{-0.23}      $\\

{\boldmath$\Sigma m_\nu^{53}$} & $< 0.912                   $ & $<0.90      $ & $<0.88     $ & $<0.86     $\\

{\boldmath$\Sigma m_\nu^{54}$} & $< 3.47                    $ & $<3.60         $ & $<3.50         $ & $<3.50         $\\

{\boldmath$\Sigma m_\nu^{61}$} & $< 2.34                    $ & $<2.40         $ & $<2.50         $ & $<2.40         $\\

{\boldmath$\Sigma m_\nu^{62}$} & $< 0.838                   $ & $<0.86      $ & $<0.79      $ & $<0.82      $\\

{\boldmath$\Sigma m_\nu^{63}$} & $< 1.33                    $ & $0.65^{+0.32}_{-0.41}      $ & $0.69\pm 0.35              $ & $0.62^{+0.29}_{-0.37}      $\\

{\boldmath$\Sigma m_\nu^{64}$} & $> 1.76                    $ & $>1.10         $ & $>1.40         $ & $>0.80         $\\
			
			\hline
			
{\boldmath$H_0                       $} & $66.49\pm 0.66             $ & $67.26\pm 0.58             $ & $66.49\pm 0.62             $ & $67.23\pm 0.56             $\\

{\boldmath$\Omega_m                  $} & $0.3298^{+0.0092}_{-0.011} $ & $0.3183\pm 0.0083          $ & $0.3298^{+0.0087}_{-0.0098}$ & $0.3188^{+0.0075}_{-0.0084}$\\

{\boldmath$\sigma_8                  $} & $0.612^{+0.073}_{-0.088}   $ & $0.554^{+0.031}_{-0.072}   $ & $0.657^{+0.13}_{-0.066}    $ & $0.594^{+0.060}_{-0.096}   $\\

{\boldmath$S_8                       $} & $0.642^{+0.078}_{-0.092}   $ & $0.570^{+0.034}_{-0.074}   $ & $0.689^{+0.13}_{-0.074}    $ & $0.612^{+0.063}_{-0.097}   $\\
			\hline
			\hline
		\end{tabular}
		\label{tab:mnuzk}}
\end{table*}

\subsection{ Scale and redshift dependent constraints on \Neff\ }

\begin{figure*}
	\centering
	\includegraphics[scale=0.5]{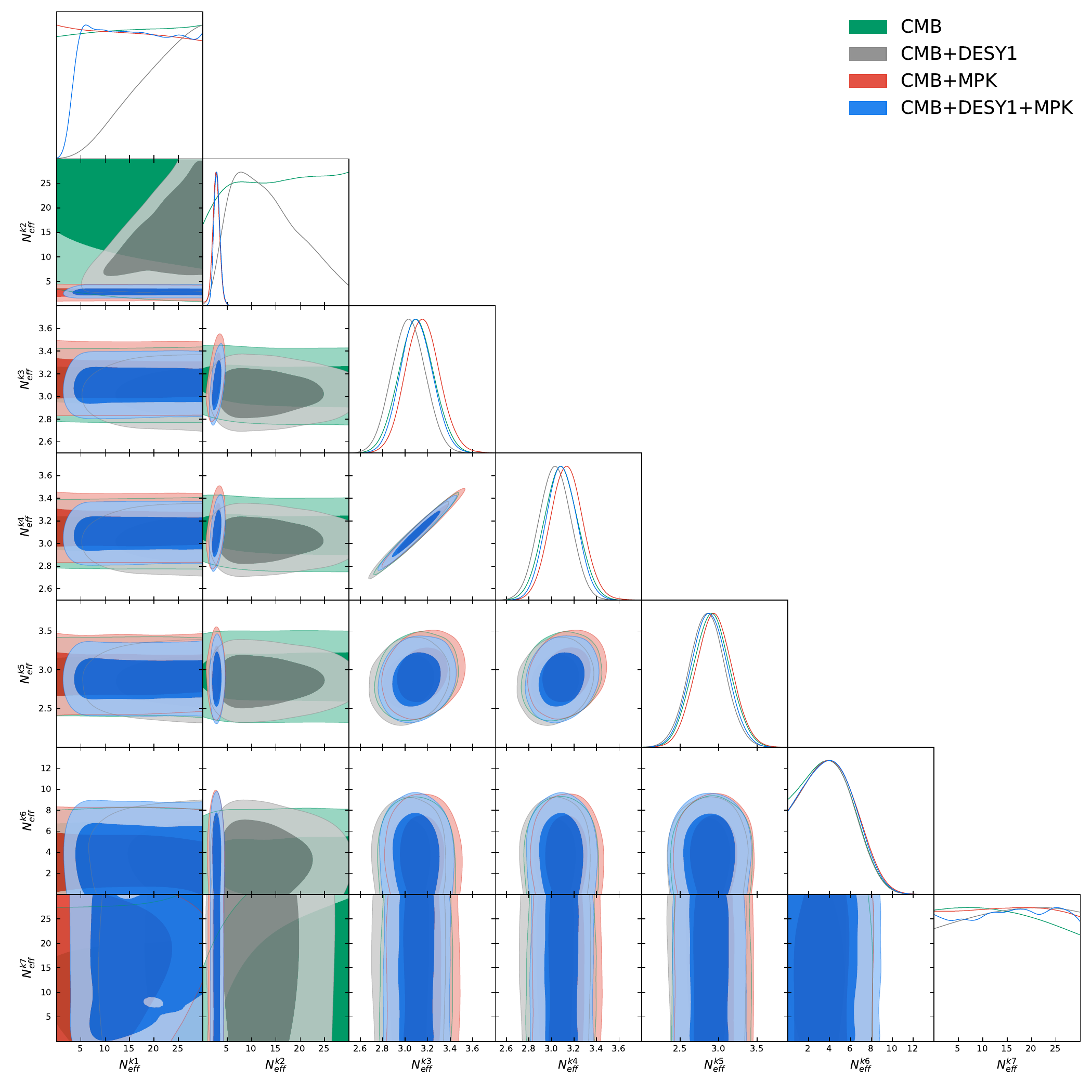}
	\caption{One-dimensional and two-dimensional marginalized constraints on the scale-dependent number of relativistic species model without massive neutrinos from CMB, CMB plus DESY1, CMB plus MPS and CMB plus DESY1 plus MPS datasets, respectively.}\label{fig:nnu_k_mnu_0}
	
\end{figure*}

\begin{figure*}
	\centering
        \includegraphics[scale=0.6]{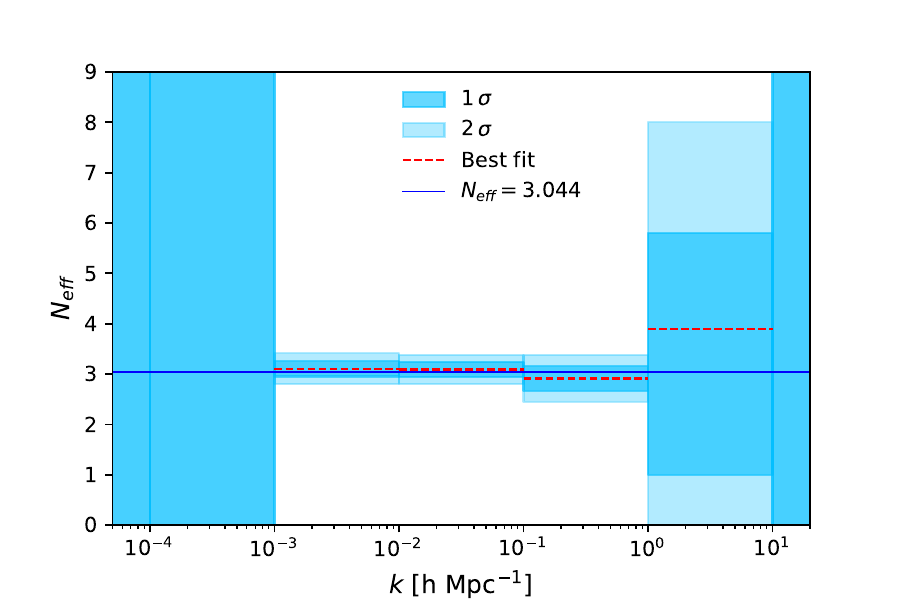}
	\caption{The tomographic reconstructions of the number of relativistic species across cosmic scales from CMB data. The blue lines denote the standard prediction of $N_{\rm eff}=3.044$, and the red dashed lines represent the best fit corresponding to the mean value of posterior distribution in each bin. }\label{fig:nnu_z_k_evolution}
\end{figure*}

Table~\ref{tab:nnuk_mnu_0} and Fig.~\ref{fig:nnu_k_mnu_0} show the constraints on \Neff\ at the different $k$-bins considered here, assuming a vanishing neutrino mass. 
At the largest scales, $[0, 10^{-4}]$~$h$/Mpc, the lack of observations leads to a completely unconstrained value of {\boldmath$N_{\rm eff}^{k7}$}. A very similar situation occurs for the $[10^{-4}, 10^{-3}]$~$h$/Mpc $k$-bin, where only a very loose upper bound of $\sim 8$ is found.
For intermediate scales, ranging from $k = 10^{-3}$~$h$/Mpc to $k = 10^{-1}$~$h$/Mpc, accurate CMB data provide very strong constraints (with additional datasets barely improving the CMB bounds), especially in the cases of {\boldmath$N_{\rm eff}^{k4}$} and {\boldmath$N_{\rm eff}^{k3}$}, where the error is slightly smaller than in the {\boldmath$N_{\rm eff}^{k5}$} case. The bound on {\boldmath$N_{\rm eff}^{k4}$} in the full dataset combination yields {\boldmath$N_{\rm eff}^{k4}$}$ = 3.09 \pm 0.14$, with an uncertainty comparable to the standard expected value without a $k$-bin analysis. Interestingly, within the $[10^{-3}, 10^{-2}]$~$h$/Mpc $k$-bin, the mean value of \Neff\ is below 3, while in the next two $k$-bins, the mean value is always above 3.
At the smallest scales, \Neff\ remains unconstrained, except for {\boldmath$N_{\rm eff}^{k2}$}, associated with the $[1, 10]$~$h$/Mpc $k$-bin, where the addition of DESY1 and/or MPS data provides limits (see associated panel in Fig.~\ref{fig:nnu_k_mnu_0}), with the mean value of \Neff\ in these cases also falling below 3.
A complete and straightforward picture of the overall analysis is provided by Fig.~\ref{fig:nnu_z_k_evolution}, where the strong constraining power of CMB data at scales between $k = 10^{-3}$~$h$/Mpc and $k = 10^{-1}$~$h$/Mpc is clearly evident.

\begin{figure*}
	\centering
	\includegraphics[scale=0.43]{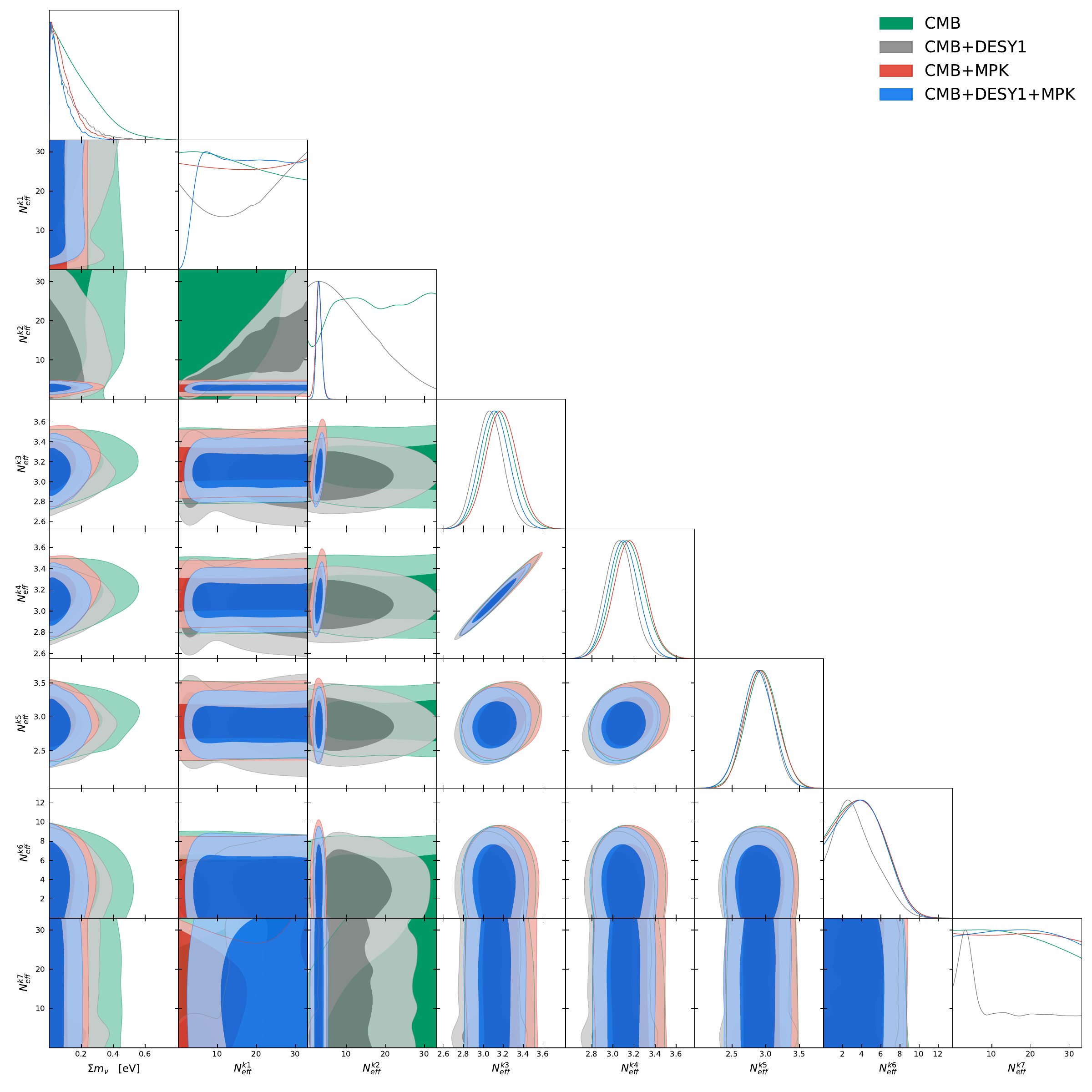}
	\caption{One-dimensional and two-dimensional marginalized constraints on the scale-dependent number of relativistic species model with free neutrino mass from CMB, CMB plus DESY1, CMB plus MPS and CMB plus DESY1 plus MPS datasets, respectively.}\label{fig:nnu_k_mnu_free}
	
\end{figure*}

Table~\ref{tab:nnuk_mnu_free} and Fig.~\ref{fig:nnu_k_mnu_free} show the analysis for different $k$ nodes in \Neff, assuming a non-vanishing neutrino mass. While at the largest scales, $[0, 10^{-4}]$~$h$/Mpc, the situation is very similar to that of massless neutrinos—leading to a completely unconstrained value of {\boldmath$N_{\rm eff}^{k7}$}—a very different scenario arises in the $[10^{-4}, 10^{-3}]$~$h$/Mpc $k$-bin, where a non-zero value for \Neff\ is found (despite the large errors). For the other $k$-bins, the results closely resemble those of the massless neutrino case, except for {\boldmath$N_{\rm eff}^{k1}$} and {\boldmath$N_{\rm eff}^{k2}$} in the CMB + DESY1 data combination, where no limits are found—unlike in the massless neutrino case.

Indeed, the tightest bound is still found for {\boldmath$N_{\rm eff}^{k4}$}$ = 3.10 \pm 0.14$ for the full dataset combination.
Concerning the neutrino mass bound, it is naturally higher than in the standard three-massive-neutrino case, due (to a minor extent) to degeneracies with \Neff\ and (to a larger extent) to the much wider neutrino parameter space. Nevertheless, the $95\%$~CL limit we find here, $\sum m_\nu < 0.205$~eV from the full dataset, is comparable to those found in minimally extended cosmologies. See, for instance, the constraints reported by Ref.~\cite{diValentino:2022njd} in extended neutrino cosmologies.

\begin{figure*}
	\centering
	\includegraphics[scale=0.4]{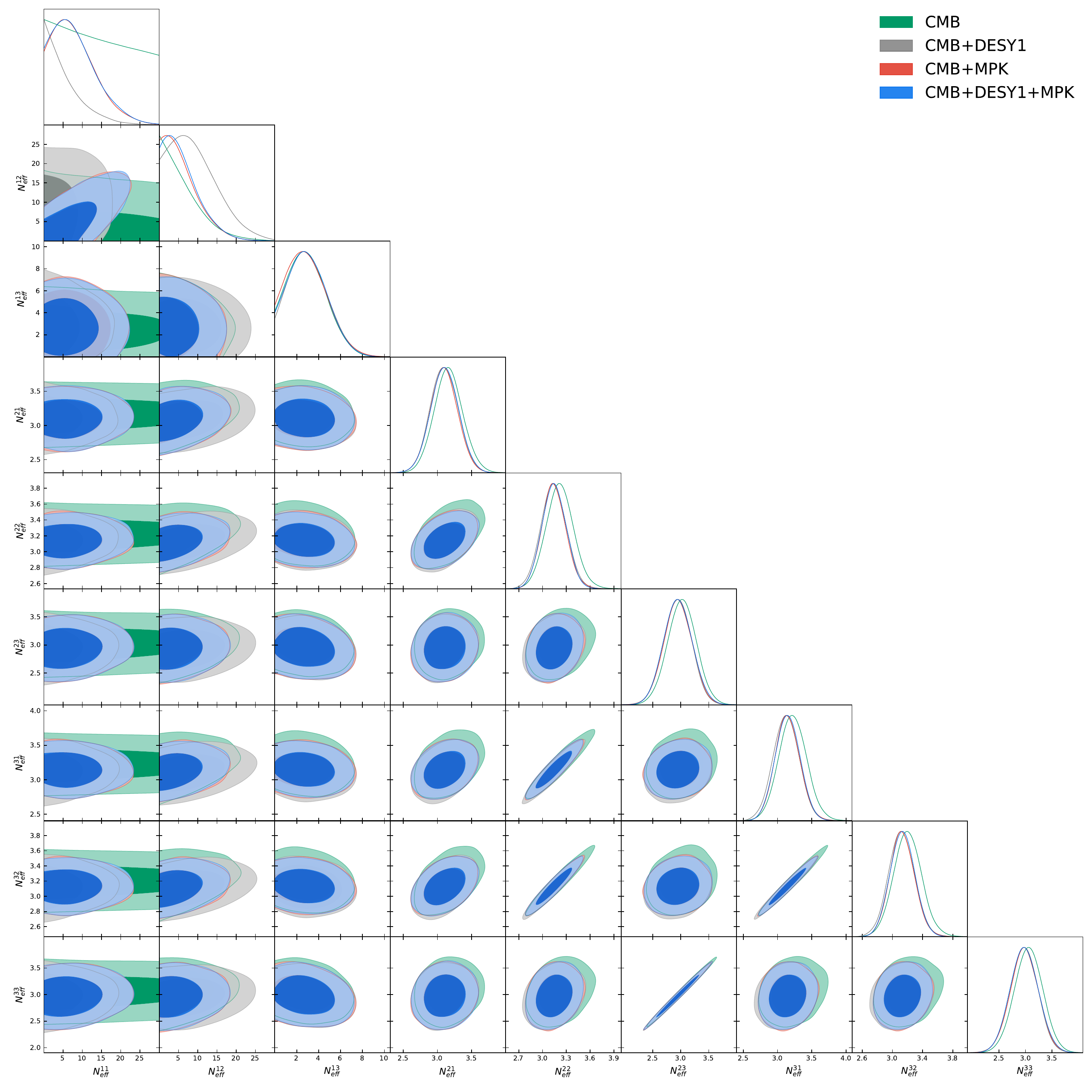}
	\caption{One-dimensional and two-dimensional marginalized constraints on the redshift- and scale-dependent number of relativistic species model from CMB, CMB plus DESY1, CMB plus MPS and CMB plus DESY1 plus MPS datasets, respectively.}\label{fig:nnuzk}
	
\end{figure*}

\begin{figure}
	\centering
	\includegraphics[scale=0.45]{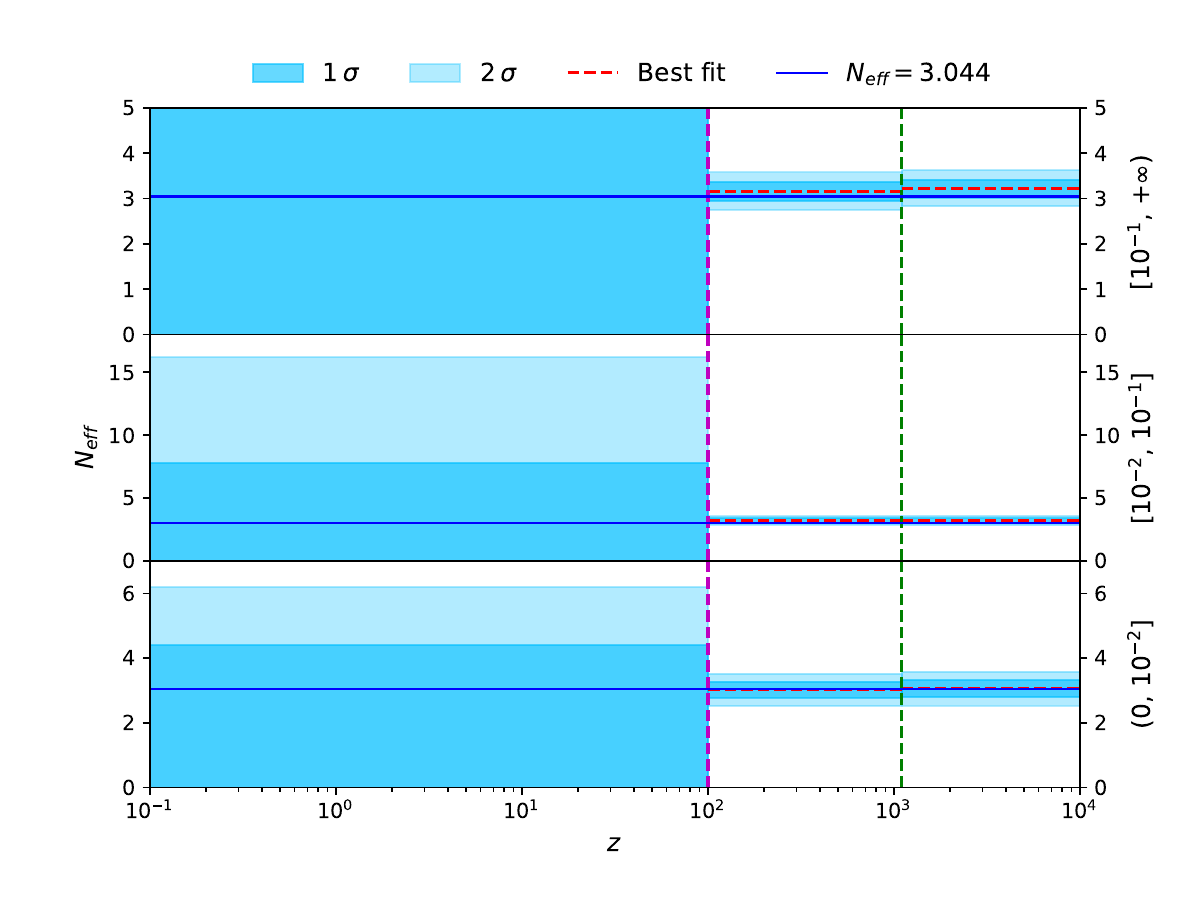}
	\caption{The tomographic reconstruction of the redshift- and scale-dependent number of relativistic species. The blue lines denote the standard prediction of $N_{\rm eff}=3.044$, and the red dashed lines represent the best fit corresponding to the mean value of posterior distribution in each bin. The magenta and green vertical dashed lines describe $z=100$ and $z=1100$, respectively.}\label{fig:nnuzkmap}
    
\end{figure}

Finally, we shall perform a modest redshift- and $k$-dependent analysis on $\Neff$. In the following, we consider three redshift bins in each $z$ and $k$ space, constraining the following parameters: {\boldmath$N_{\rm eff}^{11}$}, with $z$ in [0, 100] and $k$ in $[10^{-1}, +\infty)$~$h$/Mpc; {\boldmath$N_{\rm eff}^{12}$}, with $z$ in [0, 100] and $k$ in $[10^{-2}, 10^{-1}]$~$h$/Mpc; {\boldmath$N_{\rm eff}^{13}$}, with $z$ in [0, 100] and $k$ in $[0, 10^{-2}]$; {\boldmath$N_{\rm eff}^{21}$}, with $z$ in [100, 1100] and $k$ in $[10^{-1}, +\infty)$~$h$/Mpc; {\boldmath$N_{\rm eff}^{22}$}, with $z$ in [100, 1100] and $k$ in $[10^{-2}, 10^{-1}]$~$h$/Mpc; {\boldmath$N_{\rm eff}^{23}$}, with $z$ in [100, 1100] and $k$ in $[0, 10^{-2}]$; {\boldmath$N_{\rm eff}^{31}$}, with $z$ in [1100, $+\infty$] and $k$ in $[10^{-1}, +\infty)$~$h$/Mpc; {\boldmath$N_{\rm eff}^{32}$}, with $z$ in [1100, $+\infty$] and $k$ in $[10^{-2}, 10^{-1}]$~$h$/Mpc; and {\boldmath$N_{\rm eff}^{33}$}, with $z$ in [1100, $+\infty$] and $k$ in $[0, 10^{-2}]$~$h$/Mpc.
Table~\ref{tab:nnuzk} and Fig.~\ref{fig:nnuzk} show the results for these two-parameter bin analyses, assuming massless neutrinos. The best constraints are found in the two redshift bins [100, 1100] and [1100, $+\infty$], due to the precision of CMB data within these two redshift bins, with errors very close to those found in the standard case for $k$-bins in $[10^{-2}, 10^{-1}]$~$h$/Mpc and also $k$ within $[0, 10^{-2}]$~$h$/Mpc. The most constraining extraction of $\Neff$ is {\boldmath$N_{\rm eff}^{22}$}$ = 3.14 \pm 0.15$.
Notice also that for the largest scales, i.e., $k$ in the $[0, 10^{-2}]$~$h$/Mpc bin, the mean value of $\Neff$ is always very close to or below 3, probably suggesting a lower damping induced by $\Neff$ at these scales. For the lowest redshift bin, $\Neff$ is, in practice, unconstrained or constrained with very loose bounds. The tomographic $z$- and $k$-reconstruction of $\Neff$ is depicted in Fig.~\ref{fig:nnuzkmap}, where the high constraining power of high redshifts and intermediate scales ($[10^{-2}, 10^{-1}]$~$h$/Mpc) is clearly noticeable.

\begin{table*}[!t]
	\renewcommand\arraystretch{1.6}
	\caption{Mean values and $1\,\sigma$ ($68\%$) marginalized errors of the cosmological parameters within the scale-dependent number of relativistic species model without massive neutrinos, obtained using the CMB, CMB plus DESY1, CMB plus MPS, and CMB plus DESY1 plus MPS datasets, respectively. Note that we quote $2\,\sigma$ ($95\%$) limits for parameters that the data cannot constrain well. The symbols ``$\bigstar$'' denote parameters that remain unconstrained by the data.}
	\setlength{\tabcolsep}{5mm}{
		\begin{tabular} { l | c | c | c | c }
			\hline
			\hline
			Parameters              &  CMB & CMB+DESY1 & CMB+MPS & CMB+DESY1+MPS    \\
			\hline
			{\boldmath$\Omega_b h^2   $} & $0.02249\pm 0.00016        $ & $0.02255\pm 0.00015        $ & $0.02243\pm 0.00016        $ & $0.02251\pm 0.00015        $ \\

{\boldmath$\Omega_c h^2   $} & $0.1193\pm 0.0021          $ & $0.1177\pm 0.0018          $ & $0.1208\pm 0.0020          $ & $0.1191\pm 0.0017          $ \\

{\boldmath$100\theta_{MC} $} & $1.04101\pm 0.00038        $ & $1.04119\pm 0.00037        $ & $1.04083\pm 0.00037        $ & $1.04102\pm 0.00035        $ \\

{\boldmath$\tau           $} & $0.0519\pm 0.0080          $ & $0.0536\pm 0.0078          $ & $0.0529\pm 0.0078          $ & $0.0540\pm 0.0077          $ \\

{\boldmath${\rm{ln}}(10^{10} A_s)$} & $3.043\pm 0.024            $ & $3.035\pm 0.023            $ & $3.055\pm 0.023            $ & $3.048\pm 0.023           $ \\

{\boldmath$n_s            $} & $0.9703\pm 0.0068          $ & $0.9695\pm 0.0068          $ & $0.9697\pm 0.0069          $ & $0.9703\pm 0.0067         $ \\

{\boldmath$N_{\rm eff}^{k1}   $} & $\bigstar               $ & $<9.17             $ & $\bigstar               $ & $< 4.08        $ \\

{\boldmath$N_{\rm eff}^{k2}   $} & $\bigstar               $ & $>2.0        $ & $2.73\pm 0.75              $ & $2.92^{+0.56}_{-0.69}              $ \\

{\boldmath$N_{\rm eff}^{k3}   $} & $3.10\pm 0.15              $ & $3.03\pm 0.15              $ & $3.16\pm 0.15              $ & $3.10\pm 0.14              $ \\

{\boldmath$N_{\rm eff}^{k4}   $} & $3.09\pm 0.15              $ & $3.03\pm 0.14              $ & $3.14\pm 0.14              $ & $3.09\pm 0.14              $ \\

{\boldmath$N_{\rm eff}^{k5}   $} & $2.91\pm 0.24              $ & $2.85\pm 0.23              $ & $2.94\pm 0.23              $ & $2.88\pm 0.23              $ \\

{\boldmath$N_{\rm eff}^{k6}   $} & $< 7.84   $ & $< 7.80   $ & $< 8.05$ & $<8.50         $ \\

{\boldmath$N_{\rm eff}^{k7}   $} & $\bigstar         $ & $\bigstar         $ & $\bigstar         $ & $\bigstar         $ \\

\hline

{\boldmath$H_0                       $ } & $68.26\pm 0.87             $ & $68.93\pm 0.77             $ & $67.59\pm 0.80             $ & $68.34\pm 0.70             $ \\

{\boldmath$\Omega_m                  $ } & $0.305\pm 0.012            $ & $0.2954^{+0.0097}_{-0.0110} $ & $0.314\pm 0.012            $ & $0.3033\pm 0.0098            $ \\

{\boldmath$\sigma_8                  $ } & $0.815\pm 0.012            $ & $0.816\pm 0.012            $ & $0.816\pm 0.012            $ & $0.816\pm 0.011            $ \\

{\boldmath$S_8                  $ } & $0.821\pm 0.014            $ & $0.810\pm 0.012            $ & $0.835\pm 0.014            $ & $0.820\pm 0.012            $ \\
			
\hline
\hline
		\end{tabular}
		\label{tab:nnuk_mnu_0}}
\end{table*}

\begin{table*}[!t]
	\renewcommand\arraystretch{1.6}
	\caption{Mean values and $1\,\sigma$ ($68\%$) marginalized errors of the cosmological parameters within the scale-dependent number of relativistic species model with free neutrino mass, obtained using the CMB, CMB plus DESY1, CMB plus MPS and CMB plus DESY1 plus MPS datasets, respectively. Note that we quote $2\,\sigma$ ($95\%$) limits for parameters that the data cannot constrain well. The symbols ``$\bigstar$'' denote parameters that remain unconstrained by the data.}
	\setlength{\tabcolsep}{5mm}{
		\begin{tabular} { l | c | c | c | c }
			\hline
			\hline
			Parameters              &  CMB & CMB+DESY1 & CMB+MPS  & CMB+DESY1+MPS   \\
			\hline
			{\boldmath$\Omega_b h^2   $} & $0.02242\pm 0.00017        $ & $0.02250\pm 0.00016        $ & $0.02241\pm 0.00017        $  & $0.02249\pm 0.00016        $\\

{\boldmath$\Omega_c h^2   $} & $0.1209^{+0.0022}_{-0.0025}$ & $0.1186\pm 0.0019          $ & $0.1214\pm 0.0022          $  & $   0.1194\pm 0.0018       $\\

{\boldmath$100\theta_{MC} $} & $1.04075\pm 0.00043        $ & $1.04102\pm 0.00038        $ & $1.04074\pm 0.00039        $  & $1.04094\pm 0.00036        $\\

{\boldmath$\tau           $} & $0.0531\pm 0.0078          $ & $0.0547\pm 0.0079          $ & $0.0536\pm 0.0077          $ & $0.0551\pm 0.0078          $ \\

{\boldmath${\rm{ln}}(10^{10} A_s)$} & $3.056\pm 0.027            $ & $3.044\pm 0.024            $ & $3.061\pm 0.025            $ & $3.052\pm 0.023            $ \\

{\boldmath$n_s            $} & $0.9695\pm 0.0069          $ & $0.9690\pm 0.0067          $ & $0.9698\pm 0.0071          $ & $0.9700\pm 0.0069          $ \\

{\boldmath$\Sigma m_\nu   $} & $<0.445    $ & $<0.314    $ & $<0.242   $ & $<0.205     $ \\

{\boldmath$N_{\rm eff}^{k1}   $} & $\bigstar                   $ & $\bigstar            $ & $\bigstar                   $ & $>4.31                   $ \\

{\boldmath$N_{\rm eff}^{k2}   $} & $\bigstar           $ & $\bigstar        $ & $2.69^{+0.92}_{-0.61}      $ & $2.97^{+0.59}_{-0.69}                $ \\

{\boldmath$N_{\rm eff}^{k3}   $} & $3.15\pm 0.16              $ & $3.06\pm 0.14              $ & $3.18\pm 0.16              $ & $3.11\pm 0.15              $ \\

{\boldmath$N_{\rm eff}^{k4}   $} & $3.14\pm 0.15              $ & $3.06\pm 0.14              $ & $3.16\pm 0.15              $ & $3.10\pm 0.14               $ \\

{\boldmath$N_{\rm eff}^{k5}   $} & $2.94\pm 0.24              $ & $2.88^{+0.24}_{-0.21}      $ & $2.94\pm 0.23              $ & $2.88\pm 0.23              $ \\

{\boldmath$N_{\rm eff}^{k6}   $} & $2.90^{+3.50}_{-2.90}         $ & $2.60^{+2.90}_{-2.50}         $ & $2.90^{+3.60}_{-2.90}         $ & $2.90^{+3.60}_{-2.80}         $ \\

{\boldmath$N_{\rm eff}^{k7}   $} & $\bigstar          $ & $\bigstar          $ & $\bigstar          $ & $\bigstar          $ \\

\hline

{\boldmath$H_0         $ } & $66.10^{+2.30}_{-1.30}        $ & $67.6^{+1.60}_{-0.89}       $ & $66.6^{+1.40}_{-0.97}       $ & $67.6^{+1.10}_{-0.80}        $ \\

{\boldmath$\Omega_m                  $ } & $0.333^{+0.017}_{-0.032}   $ & $0.312^{+0.011}_{-0.021}   $ & $0.327^{+0.013}_{-0.020}   $ & $0.313^{+0.011}_{-0.016}   $ \\

{\boldmath$\sigma_8                  $ } & $0.781^{+0.036}_{-0.022}   $ & $0.795^{+0.027}_{-0.013}   $ & $0.800^{+0.020}_{-0.017}   $ & $0.802^{+0.019}_{-0.013}   $ \\

{\boldmath$S_8                  $ } & $0.822^{+0.013}_{-0.016}   $ & $0.810\pm 0.013            $ & $0.834^{+0.013}_{-0.016}   $ & $0.818\pm 0.012   $ \\
			
\hline
\hline
		\end{tabular}
		\label{tab:nnuk_mnu_free}}
\end{table*}

\begin{table*}[!t]
	\renewcommand\arraystretch{1.6}
	\caption{Mean values and $1\,\sigma$ ($68\%$) marginalized errors of the cosmological parameters within the redshift- and scale-dependent number of relativistic species model with free neutrino mass, obtained using the CMB, CMB plus DESY1, CMB plus MPS and CMB plus DESY1 plus MPS  datasets, respectively. Note that we quote $2\,\sigma$ ($95\%$) limits for parameters that the data cannot constrain well. The symbols ``$\bigstar$'' denote parameters that remain unconstrained by the data.}
	\setlength{\tabcolsep}{5mm}{
		\begin{tabular} { l | c | c | c | c  }
			\hline
			\hline
			Parameters              &  CMB & CMB+DESY1 & CMB+MPS & CMB+DESY1+MPS    \\
			\hline
			{\boldmath$\Omega_b h^2   $} & $0.02242\pm 0.00019        $ & $0.02248\pm 0.00019        $ & $0.02250\pm 0.00018        $ & $0.02250\pm 0.00019        $ \\

{\boldmath$\Omega_c h^2   $} & $0.1222\pm 0.0026          $ & $0.1203\pm 0.0023          $ & $0.1201\pm 0.0021          $ & $0.1202\pm 0.0021          $ \\

{\boldmath$100\theta_{MC} $} & $1.04068\pm 0.00042        $ & $1.04091\pm 0.00040        $ & $1.04091\pm 0.00038        $ & $1.04090\pm 0.00038        $ \\

{\boldmath$\tau           $} & $0.0526\pm 0.0080          $ & $0.0558^{+0.0074}_{-0.0083}$ & $0.0546^{+0.0071}_{-0.0081}$ & $0.0547\pm 0.0078          $ \\

{\boldmath${\rm{ln}}(10^{10} A_s)$} & $3.066\pm 0.029            $ & $3.058\pm 0.028            $ & $3.057\pm 0.026            $ & $3.066\pm 0.029             $ \\

{\boldmath$n_s            $} & $0.9691\pm 0.0096          $ & $0.9676\pm 0.0094          $ & $0.9694\pm 0.0093          $ & $0.9694\pm 0.0094          $ \\

{\boldmath$N_{\rm eff}^{11}   $} & $\bigstar            $ & $<15.40          $ & $<18.80         $ & $<19.10          $ \\

{\boldmath$N_{\rm eff}^{12}   $} & $< 16.40   $ & $< 20.20     $ & $< 14.90     $ & $< 14.80  $\\

{\boldmath$N_{\rm eff}^{13}   $} & $< 6.03   $ & $< 6.15     $ & $< 6.11        $ & $< 6.06  $ \\

{\boldmath$N_{\rm eff}^{21}   $} & $3.16\pm 0.21              $ & $3.10\pm 0.20              $ & $3.10\pm 0.20              $ & $3.11\pm 0.20             $ \\

{\boldmath$N_{\rm eff}^{22}   $} & $3.22\pm 0.18              $ & $3.14\pm 0.16              $ & $3.14\pm 0.15              $ & $3.14\pm 0.15              $ \\

{\boldmath$N_{\rm eff}^{23}   $} & $3.02\pm 0.25              $ & $2.94\pm 0.25              $ & $2.95\pm 0.25              $ & $2.95\pm 0.25              $ \\

{\boldmath$N_{\rm eff}^{31}   $} & $3.22\pm 0.20              $ & $3.13\pm 0.19              $ & $3.15\pm 0.18              $ & $3.15\pm 0.18              $ \\

{\boldmath$N_{\rm eff}^{32}   $} & $3.20\pm 0.19              $ & $3.12\pm 0.17              $ & $3.13\pm 0.16              $ & $3.13\pm 0.16              $ \\

{\boldmath$N_{\rm eff}^{33}   $} & $3.06\pm 0.27              $ & $2.97\pm 0.26              $ & $2.98\pm 0.26              $ & $2.98\pm 0.26               $ \\

\hline

{\boldmath$H_0         $ } & $67.10\pm 1.00               $ & $67.87\pm 0.96             $ & $67.93\pm 0.85             $ & $67.91\pm 0.86$ \\

{\boldmath$\Omega_m                  $ } & $0.322^{+0.014}_{-0.016}   $ & $0.310^{+0.013}_{-0.014}   $ & $0.309^{+0.011}_{-0.013}   $ & $0.310\pm 0.012   $ \\

{\boldmath$\sigma_8                  $ } & $0.797\pm 0.017            $ & $0.810\pm 0.015            $ & $0.806\pm 0.014            $ & $0.806\pm 0.014            $ \\

{\boldmath$S_8                  $ } & $0.825\pm 0.019            $ & $0.823\pm 0.014            $ & $0.818\pm 0.012            $ & $0.818\pm 0.012            $ \\
			
\hline
\hline
		\end{tabular}
		\label{tab:nnuzk}}
\end{table*}

\section{Summary}
\label{sec:conclusions}

Neutrinos are hot thermal relics with very large velocity dispersions and therefore undergo a transition from a purely relativistic component to a non-relativistic one. Given the particular nature of neutrinos, it is essential to explore the redshift- and scale-dependent limits on both neutrino masses and abundances. While redshift-dependent cosmological limits on neutrino masses and abundances have been computed previously in the literature, see e.g., Refs.~\cite{Lorenz:2021alz,Safi:2024bta}, we include here the cosmological neutrino limits versus \emph{scale} and also versus both \emph{redshift and scale}. 
We make use of a variety of cosmological observations, namely CMB, MPS, and DESY1 data. From the reconstruction of the neutrino mass as a function of scale, we find that this parameter is poorly constrained at very large scales, regardless of the datasets used in the analyses. The situation changes completely within the $[10^{-4}, 10^{-3}]$~$h$/Mpc and $[10^{-3}, 10^{-2}]$~$h$/Mpc bins, where CMB observations play a crucial role. Within the $[10^{-2}, 10^{-1}]$~$h$/Mpc $k$-bin, the CMB plus DESY1 datasets together provide the bound {\boldmath$\Sigma m_\nu^{k5}$}$ = 0.55 \pm 0.26$~eV with $68\%$~CL errors, indicating a non-zero neutrino mass with a $2\sigma$ significance level. Notice that future redshift surveys, such as Euclid and LSST, will provide unprecedented precision measurements at these scales, still within the linear regime. Therefore, the physics behind is well understood, easing the analysis of the data. Our results imply that exploiting these scales is crucial for constraining neutrino masses, or potentially find a non-zero value for those.
When focusing on redshift- and scale-dependent neutrino mass constraints, a quite interesting result is found: within the $[100, 1100]$ redshift range and the $[10^{-2}, 10^{-1}]$~$h$/Mpc $k$-bin, a non-zero neutrino mass is obtained for all data combinations, with {\boldmath$\Sigma m_\nu^{52}$}$ = 0.63^{+0.20}_{-0.24}$~eV from CMB data alone, indicating a $2$--$3\sigma$ evidence for massive neutrinos at these redshifts and scales.
This anomaly could be related to the observed deviation of the ISW amplitude, located precisely at these redshifts~\cite{Wang:2024kpu}. On the other hand, assuming a vanishing neutrino mass, a $k$-space analysis of $\Neff$ shows that at the largest scales, $[0, 10^{-4}]$~$h$/Mpc and $[10^{-4}, 10^{-3}]$~$h$/Mpc, this parameter is, in practice, unconstrained. Constraints at scales from $k = 10^{-3}$~$h$/Mpc to $k = 10^{-1}$~$h$/Mpc are found due to the high precision of CMB observations. The most constraining bound we find is {\boldmath$N_{\rm eff}^{k4}$}$ = 3.09 \pm 0.14$. This measurement is very close to the one obtained in the standard case.
At the smallest scales, $\Neff$ is unconstrained, except for {\boldmath$N_{\rm eff}^{k2}$}, associated with the $[1, 10]$~$h$/Mpc $k$-bin, where only the addition of DESY1 and/or MPS data allows for meaningful limits. For a non-vanishing neutrino mass, the situation is very similar to that of massless neutrinos, except for the $[10^{-4}, 10^{-3}]$~$h$/Mpc $k$-bin, where a mean value for $\Neff$ is found. The $95\%$~CL upper limit we find on the neutrino mass is $\sum m_\nu < 0.205$~eV from the full dataset.
Finally, we also present a joint redshift- and $k$-dependent analysis of $\Neff$, with three bins in both $z$ and $k$, i.e., 9 new parameters in total, assuming massless neutrinos. The best constraints are found in the two redshift bins [100, 1100] and [1100, $+\infty$], due to the precision of CMB data within these two redshift bins, for $k$-bins in $[10^{-2}, 10^{-1}]$~$h$/Mpc and also $k$ in $[0, 10^{-2}]$~$h$/Mpc. The most constraining extraction of $\Neff$ is {\boldmath$N_{\rm eff}^{22}$}$ = 3.14 \pm 0.15$, with errors very close to those found in the standard case.
It is very interesting to notice that, while for $N_{\rm eff}$ we can have a specific bin where the constraint is very similar to the one obtained for the standard scenario (no scale or redshift dependence), for the neutrino mass bounds this does not happen. The synergy among the different experiments at different redshifts and scales is what allows one to put a much stronger bound when $\Sigma m_\nu$ is assumed to be constant, with more than a factor 10 gain with respect to the bounds obtained on the separate bins. In this light, the results presented here serve as a guide for ongoing and future analyses of CMB and large-scale structure observations at different redshifts and scales and will be instrumental in maximizing the science output in the search for constraints on neutrino properties.

\begin{acknowledgments}
DW is supported by the CDEIGENT Fellowship of Consejo Superior de Investigaciones Científicas (CSIC).
OM acknowledges the financial support from the MCIU with funding from the European Union NextGenerationEU (PRTR-C17.I01) and Generalitat Valenciana (ASFAE/2022/020).
EDV is supported by a Royal Society Dorothy Hodgkin Research Fellowship. 
S.G.\ is supported by the Research grant TAsP (Theoretical Astroparticle Physics) funded by Istituto Nazionale di Fisica Nucleare (INFN).
This work has been supported by the Spanish MCIN/AEI/10.13039/501100011033 grants PID2020-113644GB-I00 and by the European ITN project HIDDeN (H2020-MSCA-ITN-2019/860881-HIDDeN) and SE project ASYMMETRY (HORIZON-MSCA-2021-SE-01/101086085-ASYMMETRY) and well as by the Generalitat Valenciana grants PROMETEO/2019/083 and CIPROM/2022/69. 
This article is based upon work from COST Action CA21136 Addressing observational tensions in cosmology with systematics and fundamental physics (CosmoVerse) supported by COST (European Cooperation in Science and Technology). OM acknowledges the financial support from the MCIU with funding from the European Union NextGenerationEU (PRTR-C17.I01) and Generalitat Valenciana (ASFAE/2022/020).
\end{acknowledgments}

\bibliographystyle{apsrev4-1}
\bibliography{main}

\end{document}